\documentclass[12pt]{article}     
\usepackage{amsmath,latexsym,epsfig,amssymb}

\input eqalign.sty     

\date{}
\def\ppall{\mathaccent23p}
\def\ppall{\mathaccent23p}

\def\qpall{\mathaccent23q}

\newcommand{\D}{\mathcal{D}}
\newcommand{\e}{{\rm{e}}}

\newcommand{\be}{\begin{equation}}
\newcommand{\ee}{\end{equation}}
\newcommand{\bea}{\begin{eqnarray}}
\newcommand{\eea}{\end{eqnarray}}

\newcommand{\dds}{\stackrel{\leftrightarrow}{D}}
\renewcommand{\e}{{\rm{e}}}

\begin{document}
\begin{titlepage}

\title{
{\vspace{-0cm} \normalsize
\hfill \parbox{40mm}{CERN-TH/98-251}}\\[30mm]
Moments of parton evolution probabilities on the lattice within the 
Schr\"odinger functional scheme \\ }
\author{A.\ Bucarelli$^a$, F.\ Palombi$^a$, R.\ Petronzio$^{a,b}$ and 
A.\ Shindler$^a$ \\
 {\small $^a$ Dipartimento di Fisica, Universit\`a di Roma {\em Tor Vergata}} \\ 
 {\small and INFN, Sezione di Roma II,} \\
 {\small Via della Ricerca Scientifica 1, 00133 Rome, Italy} \\
 {\small $^b$ CERN, Theory Division, CH-1211 Geneva 23, Switzerland}\\
}
\maketitle

\begin{abstract}
We define, within the Schr\"odinger functional scheme (SF), the
matrix elements of the twist-2 operators corresponding to the
first two moments of non-singlet parton densities. We perform a lattice
one-loop calculation that fixes the relation between the SF scheme and
other common schemes and shows the main source of lattice artefacts.
This calculation sets the basis for a numerical evaluation
of the non-perturbative running of parton densities.

\vspace{0.8 cm}
\noindent
\end{abstract}

\vfill
\begin{flushleft}
\begin{minipage}[t]{5. cm}
  { CERN-TH/98-251}\\

July 1998
\end{minipage}
\end{flushleft}
\end{titlepage}

\section{Introduction}

The evolution with the energy scale of hadronic structure functions can be 
computed within
renormalization group improved perturbative QCD, but the calculation of
their absolute normalization needs non-perturbative techniques.
The lattice approach is a theoretically well established framework for non-perturbative estimates
and various calculations of the first moments of hadronic structure functions
have appeared in the literature \cite{lattice_strufun}.
The results  show  a discrepancy with experimental data already 
for the non-singlet structure functions \cite{dati1,dati2}. 
There are various sources of errors that may explain the disagreement.
On the lattice side, possible systematic errors are the quenched 
approximation  and the
still rather large values of the bare coupling constant used in present 
simulations,
or equivalently the rather low values of the lattice momentum cutoff, 
which prevent a correct
estimate of the continuum limit.
On the experimental side, structure functions are reliably parametrized by 
leading-twist operators
only at large momentum transfers; a comparison at energy scales of a 
few GeV is 
affected by systematic errors due to higher-twist effects.
A standard procedure is then to evolve the lattice results from the cutoff 
scale, a couple of
GeV, to higher scales, where the comparison is safe.
The evolution is done by using perturbative results 
for the renormalization constants
of leading-twist operators on the lattice, which may be 
affected by the truncations of the
perturbative series.

Non-perturbative estimates of renormalization constants have  already been
performed  for 
local lowest-dimensional currents \cite{current} and have been seen to 
be essential for a proper definition of the
quark mass through chiral Ward identities or for disentangling 
the operator mixing
entering the evaluation of hadronic matrix elements of the effective 
four-fermion weak Hamiltonians \cite{weak_eff}.
A possible method for evaluating these renormalization constants is based on 
the Schr\"odinger
functional (SF): this method has been successfully applied to the 
study of non-perturbative
running of the coupling constant and of the quark mass \cite{alfa_mass}.
We intend to perform  a similar non perturbative estimate of the 
running of leading-twist
operators of hadronic structure functions in the SF scheme.
Non-perturbative estimates of the running of parton densities are needed
if the scale at which the hadron matrix element is renormalized is held fixed while
the continuum limit is taken. In this case the ratio of the renormalization scale over 
the lattice cutoff becomes large, and a perturbative expansion of the leading logarithms of
such a ratio is a priori not reliable.

The present paper defines the SF matrix elements of the first and second
moments of non-singlet leading-twist operators and 
reports the results on the one-loop perturbative calculation.
The calculation is useful to learn about the size of lattice artefacts 
and to fix those finite constants peculiar to the SF scheme that
can be used to make contact with more common schemes such as minimal subtraction.
In the first section we recall the basic features of the Schr\"odinger 
functional
involving fermions, in the second we discuss the choice of the matrix element 
for twist-2 non-singlet operators, and in the third we report the results of 
our calculations and some concluding remarks.

\section{The Schr\"odinger functional for fermions}

This section is only meant to recall some basic facts that have been discussed 
exhaustively in the literature \cite{sf_fond}. 
The Schr\"odinger functional represents the amplitude for the time evolution 
that takes into account quantum fluctuations of a classical
field configuration between two predetermined classical states.
It takes the form of a standard functional integral with fixed boundary 
conditions.
It has been shown that its renormalizability properties are the same as those 
of the theory in an infinite volume, modulo the possible presence of a finite 
number of boundary counterterms.
In QCD with fermions, it can be written as:

\begin{equation}
\mathcal{Z}[C^{'},\bar{\rho}',\rho';C,\bar{\rho},\rho] = \int \D[U] \D[\psi]
\D[\bar{\psi}] \rm{e}^{-S[U,\bar{\psi},\psi]}
\label{eq:sf}
\end{equation}

\noindent where $C^{'},C$ and $\bar{\rho}',\rho',\bar{\rho},\rho$ are the boundary 
values of the gauge and fermion fields respectively.
In the following discussion the classical boundary gauge field will be set 
to zero.
According to ref. \cite{campi_bordo} expectations values may involve the 
response to a variation of the classical Fermi field configurations on the 
boundaries $\zeta $:

\begin{equation}
\zeta({\bf{x}}) = \frac{\delta}{\delta\bar{\rho}({\bf{x}})}, \qquad
\bar{\zeta}({\bf{x}}) = -\frac{\delta}{\delta\rho({\bf{x}})}
\end{equation}

\begin{equation}
\zeta'({\bf{x}}) = \frac{\delta}{\delta\bar{\rho}'({\bf{x}})}, \qquad
\bar{\zeta}'({\bf{x}}) = -\frac{\delta}{\delta\rho'({\bf{x}})}
\end{equation}

\noindent as well as the fluctuating Fermi fields between the boundaries $\psi $.
In the following we will use the same notations as in ref. \cite{pert_1}.
By working out the functional integration in the presence of external sources 
one obtains the expression for the generating functional for all possible 
expectation values involving fermions, and in particular the basic non-zero 
contractions:

\begin{equation}
[\psi(x)\bar{\psi}(y)]_F = S(x,y)
\end{equation}

\begin{equation}
[\psi(x)\bar{\zeta}({\bf{y}})]_F = S(x,y)U_0(y-a\hat{0})^{-1}P_+|_{y_0=a}
\label{eq:corr1}
\end{equation}

\begin{equation}
[\psi(x)\bar{\zeta}'({\bf{y}})]_F = S(x,y)U_0(y)P_-|_{y_0=T-a}
\label{eq:corr2}
\end{equation}

\begin{equation}
[\zeta({\bf{x}})\bar{\psi}(y)]_F = P_- U_0(x-a\hat{0})S(x,y)|_{x_0=a}
\label{eq:corr3}
\end{equation}

\begin{equation}
[\zeta '({\bf{x}})\bar{\psi}(y)]_F = P_+ U_0(x)^{-1}S(x,y)|_{x_0=T-a}
\label{eq:corr4}
\end{equation}

\begin{equation}\meqalign{
[\zeta({\bf{x}})\bar{\zeta}({\bf{y}})]_F &=& P_- U_0(x-a\hat{0})S(x,y)
U_0(y-a\hat{0})P_+|_{x_0=y_0=a} \cr
&& -\frac{1}{2} P_- \gamma_k(\bigtriangledown^{\dagger}_k + 
\bigtriangledown_k)a^{-2} \delta_{{\bf{x}}{\bf{y}}}
}\end{equation}

\begin{equation}
[\zeta({\bf{x}})\bar{\zeta}'({\bf{y}})]_F = P_- U_0(x-a\hat{0})S(x,y)
U_0(y)P_- |_{x_0=a,y_0=T-a}
\end{equation}

\begin{equation}
[\zeta'({\bf{x}})\bar{\zeta}({\bf{y}})]_F = P_+ U_0(x)^{-1}S(x,y)
U_0(y-a\hat{0})^{-1}P_+ |_{x_0=T-a,y_0=a}
\end{equation}

\begin{equation}\meqalign{
[\zeta'({\bf{x}})\bar{\zeta}'({\bf{y}})]_F &=& P_+ U_0(x)^{-1}S(x,y)U_0(y)P_-
|_{x_0=y_0=T-a} \cr
&& -\frac{1}{2} P_+ \gamma_k(\bigtriangledown^{\dagger}_k + 
\bigtriangledown_k)a^{-2} \delta_{{\bf{x}}{\bf{y}}}
}\end{equation}

\noindent where $P_{\pm} = \frac{1}{2} (1 \pm \gamma_0)$ and 
$S(x,y)$ is the fermion propagator satisfying:

\begin{equation}
(D + m_{0}) S(x,y) = a^{-4}\delta_{xy}, \qquad 0<x_{0}<T.
\label{eq:propagator}
\end{equation}

The operator $D$ is the lattice version of the covariant derivative.
In this paper we will use the Wilson formulation where $D$ takes the form:

\begin{equation}
D = \frac{1}{2} \{ \gamma_{\mu} ( \bigtriangledown^{\dagger}_{\mu} + 
\bigtriangledown_{\mu} ) - a \bigtriangledown^{\dagger}_{\mu}
\bigtriangledown_{\mu} \}
\label{eq:D_wilson}
\end{equation}

To one-loop approximation, the above expression must be expanded in 
perturbation theory up to order $g^2$ by expanding the gluon unitary matrix 
$U$ in terms of vector potentials $q$:

\begin{equation}
U_\mu(x) = \{1+g a q_\mu(x) + O(g^2)\}
\label{eq:U_expand}
\end{equation}

\section{The correlations for non-singlet twist-2 operators}

In the continuum, moments of non singlet structure functions are related, 
through the operator product expansion, to hadronic matrix elements of local 
twist-2 operators of the form:

\begin{equation}
{\cal{O}}_{\mu_{1}\ldots \mu_{n}}^{qNS} = 
\bigl( \frac{i}{2}\bigl)^{n-1} {\bar{\psi}}(x)\gamma_{[\mu_{1}} 
\dds_{\mu_{2}}\cdots \dds_{\mu_{n}]} 
\frac{\lambda^f}{2} \psi(x)\ +\ \mbox{trace terms}
\label{eq:twist two_continuum}
\end{equation}

\noindent where $\dds_{\mu}$ is the covariant derivative, 
[``indices''] means symmetrization, and $p$ is the hadron momentum 
that can be assumed light-like
up to power corrections of order $M^2/Q^2$ with $Q$ the four-momentum
 transfer of the vector boson probing the hadron structure.

The twist is defined from the difference between the engineering dimensions 
of the operator and its angular momentum. 
Indeed, all listed operators belong to irreducible representations 
of the angular momentum. 

On the lattice, the discretization of the covariant derivative can be done in 
a standard way:

\begin{equation}\meqalign{
\bigtriangledown_{\mu} \psi(x) &=& 
\frac{1}{a}[U_\mu(x) \psi(x+a\hat{\mu}) - \psi(x)] \cr
\bigtriangledown^{\dagger}_{\mu} \psi(x) &=& 
\frac{1}{a}[\psi(x)-U_\mu(x-a\hat{\mu})^{-1} 
\psi(x-a\hat{\mu})]
\label{eq:D_discrete}
}\end{equation}

\noindent but in general, owing to the lower (hypercubic) symmetry of the Euclidean 
lattice action with respect to that of the continuum ( all 4-d rotations), 
the identification of a given irreducible representation may require some 
particular combinations of operators. 
This classification has been discussed for
example in refs. \cite{h4_first} and \cite{h4_second}. 
With the Schr\"odinger functional the symmetry 
is in general further reduced, from hypercubic to cubic, because of the fixed 
boundary conditions in time that mark this direction with respect 
to the others. 
In this paper, we will concentrate on the calculation of the 
first two moments to which we associate the following irreducible operators:

\begin{equation}\meqalign{
O^q_{12} &=& \frac{1}{4}\bar\psi \gamma_{[1} \dds_{2]}\frac{\lambda^f}{2}\psi \cr
O^q_{123} &=& \frac{1}{8}\bar\psi \gamma_{[1} 
\dds_{2} \dds_{3]}\frac{\lambda^f}{2}\psi .
\label{eq:lat_operators}
}\end{equation}

\noindent where $\lambda^f$ is a flavour matrix.
\noindent These are a subset of the basis described in ref. \cite{h4_second}, 
involving only spatial indices and are multiplicatively renormalizable.
We define the  matrix elements of the first two moments by the 
observables:

\begin{equation}\meqalign{
f_{0_{12}}(x_0) &=& f_2(x_0) = -a^6\sum_{\bf{y},\bf{z}} \rm{e}^
{i\bf{p}(\bf{y}-\bf{z})}
\langle \frac{1}{4} \bar\psi(x) \gamma_{[1} 
\dds_{2]}\frac{1}{2} \tau^3 \psi(x) 
\bar\zeta({\bf{y}}) \Gamma \frac{1}{2} \tau^3 \zeta({\bf{z}})\rangle \cr
f_{0_{123}}(x_0) &=& f_3(x_0) = -a^6\sum_{\bf{y},\bf{z}} \rm{e}^
{i\bf{p}(\bf{y}-\bf{z})}
\langle \frac{1}{8}\bar\psi(x) \gamma_{[1} \dds_{2}
\dds_{3]}\frac{1}{2}\tau^3 \psi(x)
\bar\zeta({\bf{y}}) \frac{1}{2} \Gamma \tau^3 \zeta({\bf{z}})\rangle
\label{eq:S_observable}
}\end{equation}

\noindent where the contraction of the classical fields is non-vanishing if the matrix 
$\Gamma $ satisfies: $\Gamma P_{-(+)} = P_{+(-)}$ and $p$ is the 
momentum of the classical field sitting on the boundary. They can be seen as the
operator matrix elements between the vacuum and ``$\rho$''-like classical state sitting
at the $T=0$ boundary.

We take the limit of massless quarks, which in the numerical simulations
can be monitored  via axial Ward identities. In the
SF framework it is possible to work at zero physical quark mass
because a natural infrared cutoff to the Dirac operator eigenmodes is
provided by the time extent of the lattice. 
This choice simplifies the recursive procedure at finite volume that 
leads to the reconstruction of the continuum non-perturbative
 renormalization constant of the operators.

The matrix element of the operator for the first moment involves 
two directions and three for the second moment. 
These directions must be provided by external vectors:
we have chosen to obtain one of them from the contraction matrix $\Gamma $,
i.e. from the polarization of the vector classical state:

\begin{equation}
\Gamma = \gamma_k, \qquad k=1,2,3 
\label{eq:gamma}
\end{equation}


\noindent and the remaining ones from the momentum $p$ of the classical Fermi 
field at the boundary.
This choice gives for the tree level a non-vanishing matrix element 
in the massless quark limit, where we evaluate our correlations. 
The tree-level correlation can be easily calculated
and reads, for the first moment:

\begin{equation}
f^{(0)}_2(x_0) = \frac{i\ppall_1 N}{R(p)^2} \left[ (-i\ppall_0) 
\left(M_-(p) {\rm{e}}^{-2\omega({\bf{p}})x_0} - 
M_+(p) {\rm{e}}^{-2\omega({\bf{p}})(2T-x_0)}\right) 
\right]
\label{eq:tree}
\end{equation}

\noindent where 

\begin{equation}
\hat p_\mu = (2/a) \sin (ap_\mu/2), \qquad \ppall_\mu = (1/a) \sin (ap_\mu)
\end{equation}

\begin{equation}
M(p) = m + \frac{1}{2}a \hat{p}^2, \qquad M_{\pm} = M(p) \pm i \ppall_0
\end{equation}

\begin{equation}
\textrm{sinh } \left[\frac{a}{2}\omega({\bf{q}})\right] = 
\frac{a}{2} \left\{ \frac{ {\bf{\qpall}} ^2 + 
(m +\frac{1}{2} a\hat{{\bf{q}}}^2)^2}
{1+a (m +\frac{1}{2} a\hat{{\bf{q}}}^2)} \right\}^{\frac{1}{2}}
\end{equation}

\begin{equation}
R(p) = M(p) 
\left\{1-\textrm{e}^{-2\omega({\bf{p}})T}\right\} -
i\ppall_0 \left\{1+\textrm{e}^{-2\omega({\bf{p}})T}\right\}.
\end{equation}

We have chosen for convention 
$\Gamma = \gamma_{2}$ and $\bold{p} = (p1,0,0)$.
In the continuum and in the massless limit this expression reduces to:

\begin{equation}
f^{(0)}_2(x_0) = \frac{N p_1}{(1+\e^{-2 p_1 T})^2}
\left[\e^{-2 p_1 x_0} + \e^{-2 p_1 (2T-x_0)} \right].
\label{eq:tree_continuum}
\end{equation}

Notice that the symmetric choice 
$\Gamma = \gamma_{2} p2 + \gamma_{1} p1$ and $\bold{p} = (p1,p2,0)$
would lead to:

\begin{equation}\meqalign{
f^{(0)}_2(x_0) &=& \frac{\ppall_1^3 \ppall_2 N}{R(p)^2} 
\left( 4 \e^{-2 \omega({\bf{p}}) T} \right) +
\frac{\ppall_1 \ppall_2^3 N}{R(p)^2} 
\left( 4 \e^{-2 \omega({\bf{p}}) T} \right) \cr
&+& \frac{\ppall_1 \ppall_2 M(p)^2 N}{R(p)^2}
\left( 2 \e^{-2 \omega({\bf{p}}) x_0} + 
2 \e^{-2 \omega({\bf{p}}) (2T - x_0)} \right) \cr
&+& \frac{\ppall_1 \ppall_2 M(p) (-i \ppall_0) N}{R(p)^2}
\left( 2 \e^{-2 \omega({\bf{p}}) x_0} -
2 \e^{-2 \omega({\bf{p}}) (2T - x_0)} \right)
\label{eq:tree_symm}
}\end{equation}

which in the massless continuum limit reduces to

\begin{equation}
f^{(0)}_2(x_0) = \frac{p_1^3 p_2 N}{(1+\e^{-2 |{\bf{p}}| T})^2} 
\left( 4 \e^{-2 |{\bf{p}}| T} \right)  
+ \frac{p_1 p_2^3 N}{(1+\e^{-2 |{\bf{p}}| T})^2} 
\left( 4 \e^{-2 |{\bf{p}}| T} \right)
\label{eq:tree_symm_continuum}
\end{equation}

In the massless continuum limit, only the contribution proportional to 
$\exp (-2 \omega T)$  survives, while at finite lattice spacing and for $x_0$ smaller than 
T the dominant term is $\exp (-2 \omega x_0)$,
i.e.  in this case the signal is dominated by lattice artefacts.
Another advantage of using the non symmetric contraction is the economy 
in the number of non-zero momentum components. 

The quantization of momenta on a finite lattice is one of the
major sources of lattice artefacts. Indeed, with ordinary periodic boundary
conditions the momentum cannot be chosen, in each direction, smaller 
than a minimum value proportional to the inverse lattice size. 
The presence of more independent momentum components increases the 
absolute value of the total momentum and therefore of the
lattice artefacts associated to its size. 
In this respect, we will see that a different choice of boundary
conditions can provide a ``finite-size'' momentum not submitted to 
quantization.

For the second moment, we choose 
$\Gamma = \gamma_{3}$ and $\bold{p} = (p1,p2,0)$.
The tree-level expression in this case is just proportional to the one for the 
first moment:

\begin{equation}
f^{(0)}_3(x_0) = \frac{\ppall_1 \ppall_2 N}{R(p)^2} 
\left[ (i\ppall_0) \left(M_-(p) {\rm{e}}^{-2\omega({\bf{p}})x_0} - 
M_+(p) {\rm{e}}^{-2\omega({\bf{p}})(2T-x_0)}\right) \right] = 
i \ppall_2 f^{(0)}_2(x_0)
\label{eq:tree_second_proportional}
\end{equation}

The observable is taken as the ratio of the correlation 
at one loop 
normalized by its tree-level expression and must be expurgated from the 
renormalization constant of the classical boundary fields $\zeta$. 
Following ref. \cite{pert_2}, this is achieved through the
normalization by the quantity called $f_1$, also normalized by its tree-level 
expression.

Finally, some care should be taken so as to ensure the massless quark 
limit at order $g^2$.
The breaking of chiral simmetry of the Wilson action entails a non-zero shift
of the quark mass from the naive value at order $g^2$:

\begin{equation}
m_c^{(1)} = -\frac{2}{3} \int_{-\pi}^{\pi} \frac{d^4 q}{2\pi^4}
\left\{\frac{1}{\hat{q}^2} \left[ 3 + 
\frac{\hat{q}^2(4-\frac{1}{4}\hat{q}^2)-\qpall^2}
{\qpall^2 + \frac{1}{4}(\hat{q}^2)^2} \right] \right\}.
\label{eq:mass shift}
\end{equation}

Including the mass shift simply amounts to a further contribution 
coming from the
derivative of the tree-level expression calculated at the zero naive quark 
mass value times the coefficient of the order $g^2$ mass shift.

\section{The results of the calculation}

The expression for the observable at one loop in the continuum can be 
parametrized as:

\begin{equation}\meqalign{
& Z(pL,x_0/L,a/L) = 1 + g^2 Z^{(1)}(a/L) \cr 
& \noindent {\rm{with}} \qquad Z^{(1)}(a/L) = b_0 + c_0 \ln(a/L) + \sum_{k=1}^{\infty}
a^k \frac{b_k + c_k \ln(a/L)}{L^k},
\label{eq:res_parametrisation}
}\end{equation}

\noindent where the scaling variables  $pL$ and  $x_0/L$ are kept constant when
the number of lattice points is sent to infinity. 
Indeed, beyond the free field 
case, this limit does not exist because of the ultraviolet divergence
manifested through the logarithmic term that breaks the free-field theory 
scale invariance. In the following, the momentum will be set to its minimum value,
i.e. $pL = 2 \pi$.

The operator needs renormalization to be finite in the continuum limit:
for example one can fix the operator 
matrix element equal to its tree-level value at $\mu = 1/L$.
At one loop one gets:

\begin{equation}\meqalign{
O^{R}(\mu) &=& (1 - g^2 Z^{(1)}(a \mu))O^{bare}(a) \cr
&=& (1 - g^2 Z^{(1)}(a \mu))(1 + g^2 Z^{(1)}(a/L)) O^{tree}.
\label{eq:ren_condition}
}\end{equation}

The coefficients $b_0,c_0$ are the interesting ones in the 
continuum limit, but at finite 
lattice spacing, i.e. with a finite number of lattice points, 
the correlation also contains corrections that decrease with the 
inverse powers of the number of points $N$. 
Our results are for the ordinary Wilson action without any improvement, and 
lattice artefacts start at order $1/N$ for both coefficients.

General formulae for calculating the perturbative expansion within the SF 
can be found in refs. \cite{pert_2, pert_1}, where the correlation in eqs.
(\ref{eq:corr1} to \ref{eq:corr4}) is calculated as:

\begin{equation}\meqalign
[\psi(x)\bar{\zeta}({\bf{y}})]_F = \frac{\delta\psi_{\rm{cl}}(x)}
{\delta\rho({\bf{y}})} \cr
[\psi(x)\bar{\zeta}'({\bf{y}})]_F = \frac{\delta\psi_{\rm{cl}}(x)}
{\delta\rho'({\bf{y}})} \cr
[\zeta({\bf{x}})\bar{\psi}(y)]_F = \frac{\delta\bar{\psi}_{\rm{cl}}(y)}
{\delta\bar{\rho}({\bf{x}})} \cr
[\zeta '({\bf{x}})\bar{\psi}(y)]_F = \frac{\delta\bar{\psi}_{\rm{cl}}(x)}
{\delta\bar{\rho}'({\bf{y}})}
\label{eq:psizeta_psiclassical}
\end{equation}

\noindent and in turn 

\begin{equation}\meqalign{
\psi_{\rm{cl}}(x) &=& a^3 \sum_{\bf{y}} \left\{ S(x,y)U(y - a\hat{0},0)^{-1} 
\rho({\bf{y}})|_{y_0 = a} \right. \cr
&+& \left. S(x,y) U(y,0) \rho'({\bf{y}})|_{y_0 = T-a} \right\}
\label{eq:psicl}
}\end{equation}

We have done the calculation independently by using the notion of 
$\psi_{\rm{cl}}$ 
and the explicit expressions 
of ref. \cite{pert_1} or
by using directly the correlations given by eqs.
\ref{eq:corr1} to \ref{eq:corr4} 
and the perturbative expansion of the Dirac propagator. 
In the latter case we obtain also diagrams where the gluon field 
emerges from the time links connecting the Dirac propagator to the zero 
time slice.
In both cases the calculation was done in a time-momentum representation. 
The algebraic manipulations leading to the finite sums performed 
numerically were done with the algebraic program ``FORM''.

The final Fortran result was obtained with double-precision accuracy.
We have run all even lattice sizes ranging from 8 to 36 in the case of the 
first moment
and to 40 for the second moment. The latter choice was forced by the slower
approach to the continuum for the second moment.
The fit to the $N$ dependence of our results has been made by the expression

\begin{equation}
Z^{(1)}(N) = B_0 + C_0 \ln(N) + \sum_{k = 1, 2} 
\frac{B_k + C_k \ln(N)}{N^{k}} \quad, {\rm{with}} \quad N = L/a,
\label{eq:fittype}
\end{equation}

\noindent excluding higher-order terms.
The stability of the fit was checked against a change in the value of the 
smallest volume included. 
After checking the agreement to the per cent level with the expected value  
of the coefficient of the logarithm $C_0$, we
fix it to its theoretical value and determine the remaining parameters.
As an example, we report in Table $1$ the results for the 
first moment and for the correlations at $x_0= L/2$ of the constant $B_0$.

\begin{table}
\begin{center}
\begin{tabular}{|@{\qquad} c @{\qquad}|@{\qquad} c @{\qquad}|
@{\qquad} c @{\qquad}|@{\qquad} c @{\qquad}|}
\hline
Range & Parameters & Constant & $\chi^2$\\
\hline 
$ 24$--$34 $ & $ 5 $ & $ 0.0971 $ & $ 10^{-2} $ \\ 
\hline 
$ 22$--$34 $ & $ 5 $ & $ 0.0968 $ & $ 10^{-3} $ \\ 
\hline 
$ 20$--$34 $ & $ 5 $ & $ 0.0964 $ & $ 10^{-5} $ \\ 
\hline 
\end{tabular}
\caption{\footnotesize{Values of the constant of the correlation 
function of the first moment with $x_0 = L/2$.}}
\end{center}
\label{tab:fit1}
\end{table}

The values are shown for different fitting intervals and, in the last column, 
we report the $\chi^2$ values with a reference error of $10^{-5}$ on the theoretical 
points .
An alternative determination can be  achieved by making suitable 
combinations of the data
where some of the dominant $1/L$ and $1/L^2$ corrections are eliminated.
This leads to the results of Table 2, where we indicate 
the type  of fitting coefficients eliminated from the general eq. and 
the fitting interval used.

Our final results for the constants, for the first and second moment, 
 are in Table \ref{tab:constants}.

\begin{table}
\begin{center}
\begin{tabular}{|@{\qquad} c @{\qquad}|@{\qquad} c @{\qquad}|
@{\qquad} c @{\qquad}|@{\qquad} c @{\qquad}|}
\hline
Range & Eliminated parameters & Constant & $\chi^2$\\
\hline 
$ 24$--$30 $ & $ \ln (N)/N, 1/N, 1/N^2 $ & $ 0.0967 $ & $ 1.5 $ \\ 
\hline 
$ 24$--$30 $ & $ \ln (N)/N, 1/N $ & $ 0.0967 $ & $ 10^{-2} $ \\ 
\hline 
\end{tabular}
\caption{\footnotesize{
Values of the constant of the correlation function of the first moment with $x_0 = L/2$ 
after elimination of some subleading contributions.}}
\end{center}
\label{tab:fit2}
\end{table}

\begin{table}
\begin{center}
\begin{tabular}{|@{\qquad} c @{\qquad}|@{\qquad} c @{\qquad}|
@{\qquad} c @{\qquad}|}
\hline
Moment & Definition & Constant \\
\hline
First & $ x_0 = L/4 $ & $B_0 = 0.084(1)$  \\
\hline
First  & $ x_0 = L/2 $ & $B_0 = 0.0967(5)$  \\
\hline
Second & $ x_0 = L/4 $ & $B_0 = 0.010(5)$  \\
\hline
Second  & $ x_0 = L/2 $ & $B_0 = 0.008(2)$  \\
\hline
\end{tabular}
\caption{\footnotesize{
Values of the constants of the correlation functions $f_2$ and $f_3$ 
 with the operator insertion at $x_0 = L/2$ and $x_0 = L/4$.}}
\label{tab:constants}
\end{center}
\end{table}

\begin{figure}[ht]
\hspace{0cm}
\vspace{-0.0cm}
\centerline{\psfig{figure=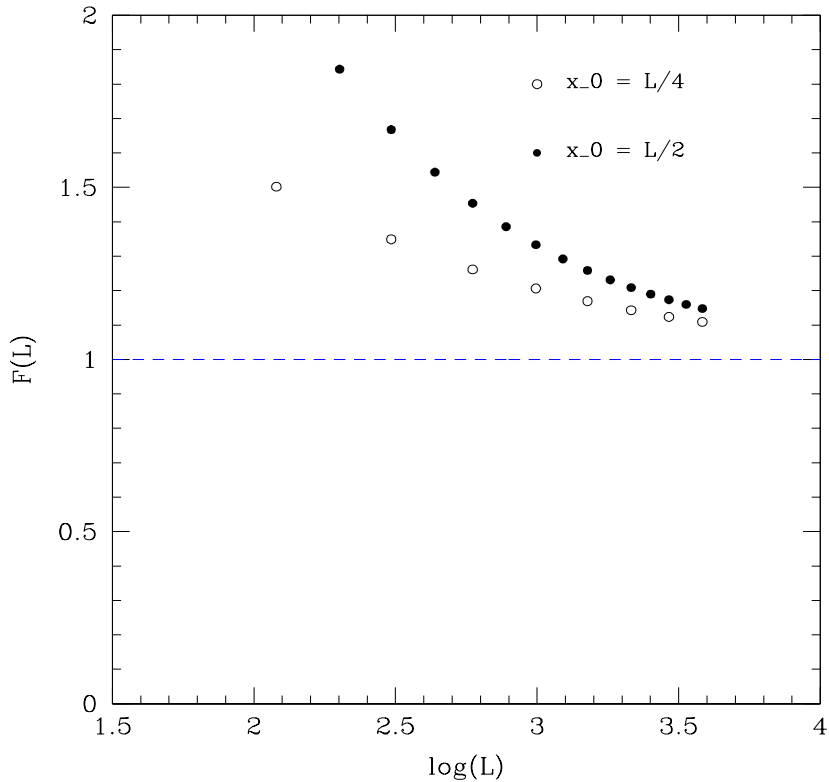,height=4.3in,width=4.5in}}
\vspace{-0.0cm}
\caption{\footnotesize{
The ratio of perturbative finite lattice 
renormalization constant with their continuum value for the first moment 
in the two cases of $x_0 = L/4 (\circ)$ and $x_0 = L/2 (\bullet)$.}
\label{fig:ratio1}}
\end{figure}

\begin{figure}[ht]
\hspace{0cm}
\vspace{-0.0cm}
\centerline{\psfig{figure=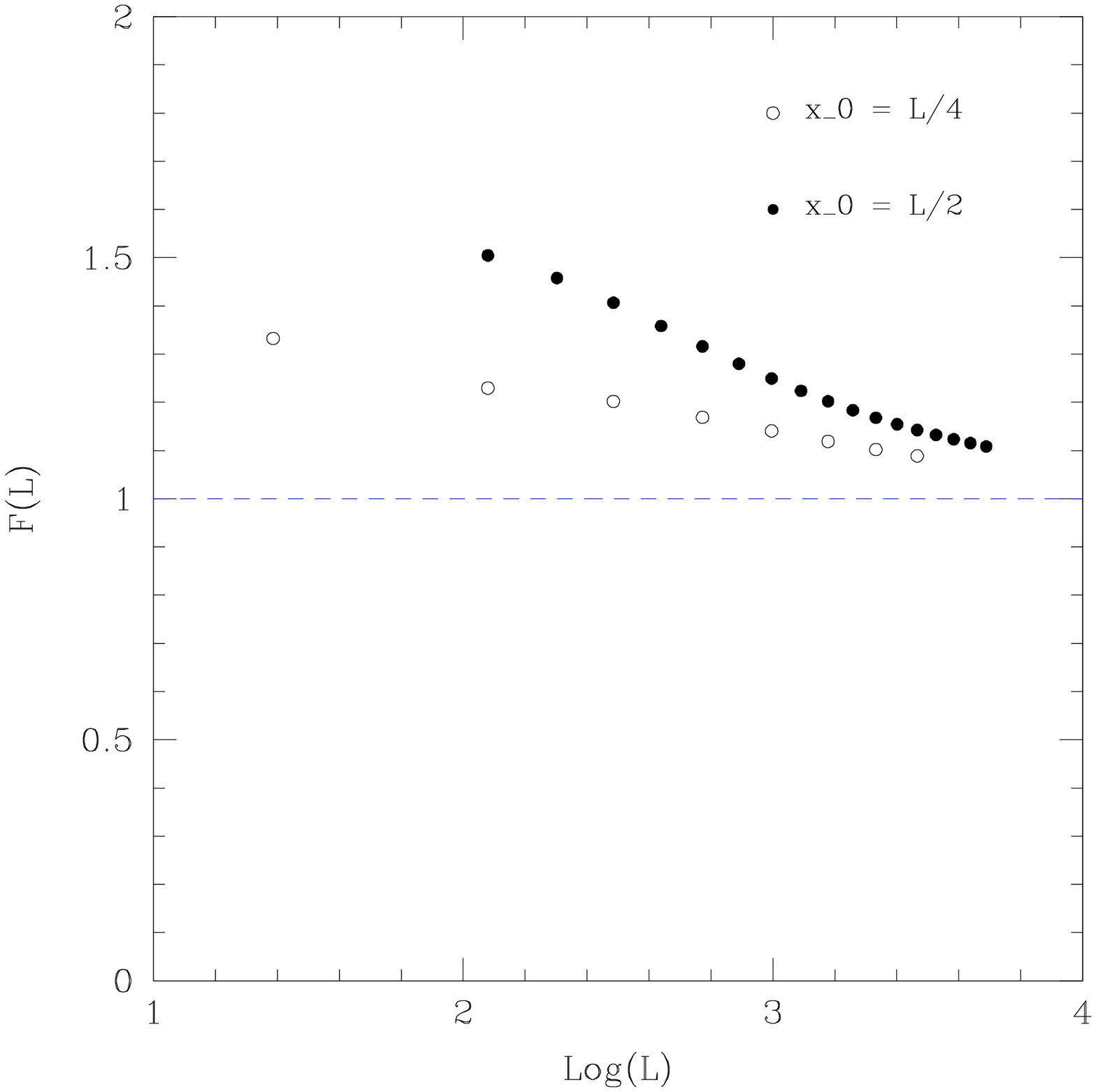,height=4.3in,width=4.5in}}
\vspace{-0.0cm}
\caption{\footnotesize{
The ratio of perturbative finite lattice 
renormalization constants with their continuum value for the second moment 
in the two cases of $x_0 = L/4 (\circ)$ and $x_0 = L/2 (\bullet)$.}
\label{fig:ratio2}}
\end{figure}

The approach to the continuum values is very slow, as can be seen from 
figures \ref{fig:ratio1} and \ref{fig:ratio2} where we 
compare the ratio of perturbative finite lattice coefficients with their 
continuum values for the two moments and the two $x_0$ cases.

In the case of the second moment the slope of the approach is lower, 
indicating that momentum quantization is an important source of lattice 
artefacts. 
The continuum is approached from above in
all cases, similarly to what happens for the proton matrix 
elements, by comparing the standard Wilson action with a partially 
non-perturbative $O(a)$ improvement \cite{dati2}.

The use of a physical momentum is not the only way to obtain non-vanishing
matrix elements for the operators that we have considered. Indeed,
the choice of periodic boundary conditions for fermions different from the 
usual ones \cite{priv_comm}:

\begin{equation}
\psi(x) = \psi( x + L \hat{k} )
\label{eq:oldboundarycond}
\end{equation}

\noindent to:

\begin{equation}
\tilde{\psi}(x) = \e^{i \theta_k} \tilde{\psi}( x + L \hat{k} )
\label{eq:newboundarycond}
\end{equation}

\noindent where $\theta$ is an arbitrary constant phase, introduces a 
``finite-size momentum $\theta$'' probed
by the operator only when it involves fields on the two sides of the boundary
that can play the same role as the ordinary momentum.
One can also distribute the phase to all lattice points 
by an Abelian transformation on the Fermi fields:

\begin{equation}
\tilde{\psi}(x) = \e^{i \frac{\theta}{L} x} \psi(x),
\label{eq:abeliantrans}
\end{equation}

\noindent which changes the form of the lattice derivative from

\begin{equation}\meqalign{
\bigtriangledown_{\mu} \tilde{\psi}(x) &=& 
\frac{1}{a}[U_\mu(x) \tilde{\psi}(x+a\hat{\mu}) - \tilde{\psi}(x)] \cr
\bigtriangledown^{\dagger}_{\mu} \tilde{\psi}(x) &=&
\frac{1}{a}[\tilde{\psi}(x) - U_\mu(x-a\hat{\mu})^{-1} 
\tilde{\psi}(x-a\hat{\mu})]
\label{eq:latticederiv_old}
}\end{equation}

\noindent to:

\begin{equation}\meqalign{
\bigtriangledown_{\mu} \psi(x) &=& 
\frac{1}{a}[\lambda_\mu U_\mu(x) \psi(x+a\hat{\mu}) - \psi(x)] \cr
\bigtriangledown^{\dagger}_{\mu} \psi(x) &=& 
\frac{1}{a}[\psi(x)-\lambda_\mu^{-1} U_\mu(x-a\hat{\mu})^{-1} 
\psi(x-a\hat{\mu})],
\label{eq:latticederiv_new}
}\end{equation}

\noindent where 

\begin{equation}
\lambda_\mu = e^{ia\theta_\mu / L},  \qquad \theta_0 = 0,
\qquad -\pi < \theta_k \leq \pi.
\end{equation}

In this case the operator feels a ``finite-size momentum $\theta/L$''at 
all lattice points.
By taking the average of the operator insertion over all points, 
the same result is recovered.

We have repeated the perturbative calculation at zero momentum and non-zero
``finite size momentum'' 
$\theta = 0.1/L$, to be compared with the minimal lattice momentum 
$p_{min} = 2 \pi/L$.

\begin{figure}[ht]
\hspace{0cm}
\vspace{-0.0cm}
\centerline{\psfig{figure=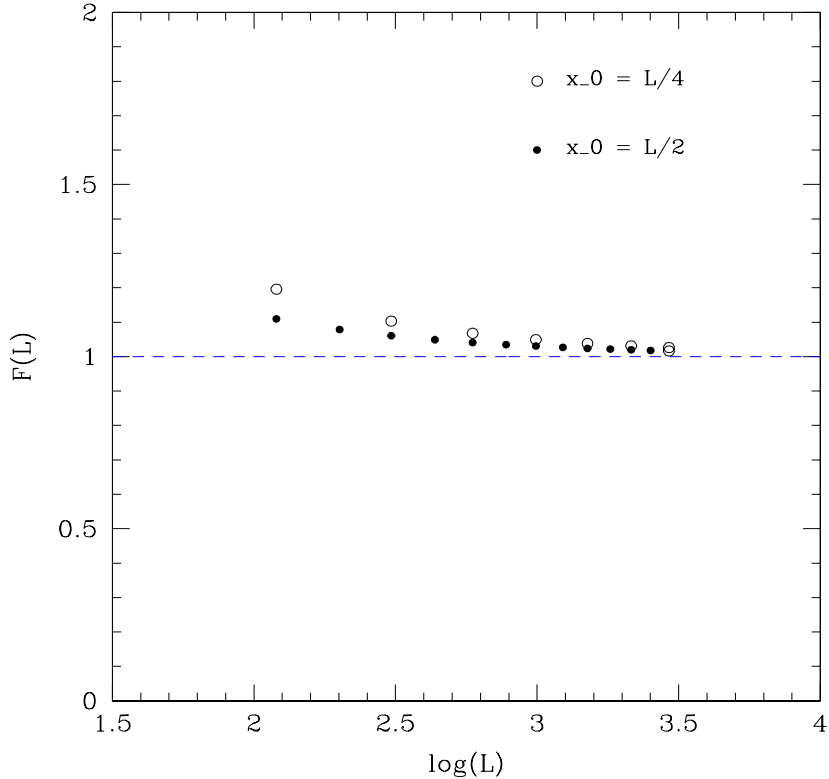,height=4.3in,width=4.5in}}
\vspace{-0.0cm}
\caption{\footnotesize{
The ratio of perturbative finite lattice 
renormalization constants with their continuum value for the first moment 
in the two cases of $x_0 = L/4 (\circ)$ and $x_0 = L/2 (\bullet)$ using 
``finite size momentum''.}
\label{fig:rate1}}
\end{figure}

\begin{figure}[ht]
\hspace{0cm}
\vspace{-0.0cm}
\centerline{\psfig{figure=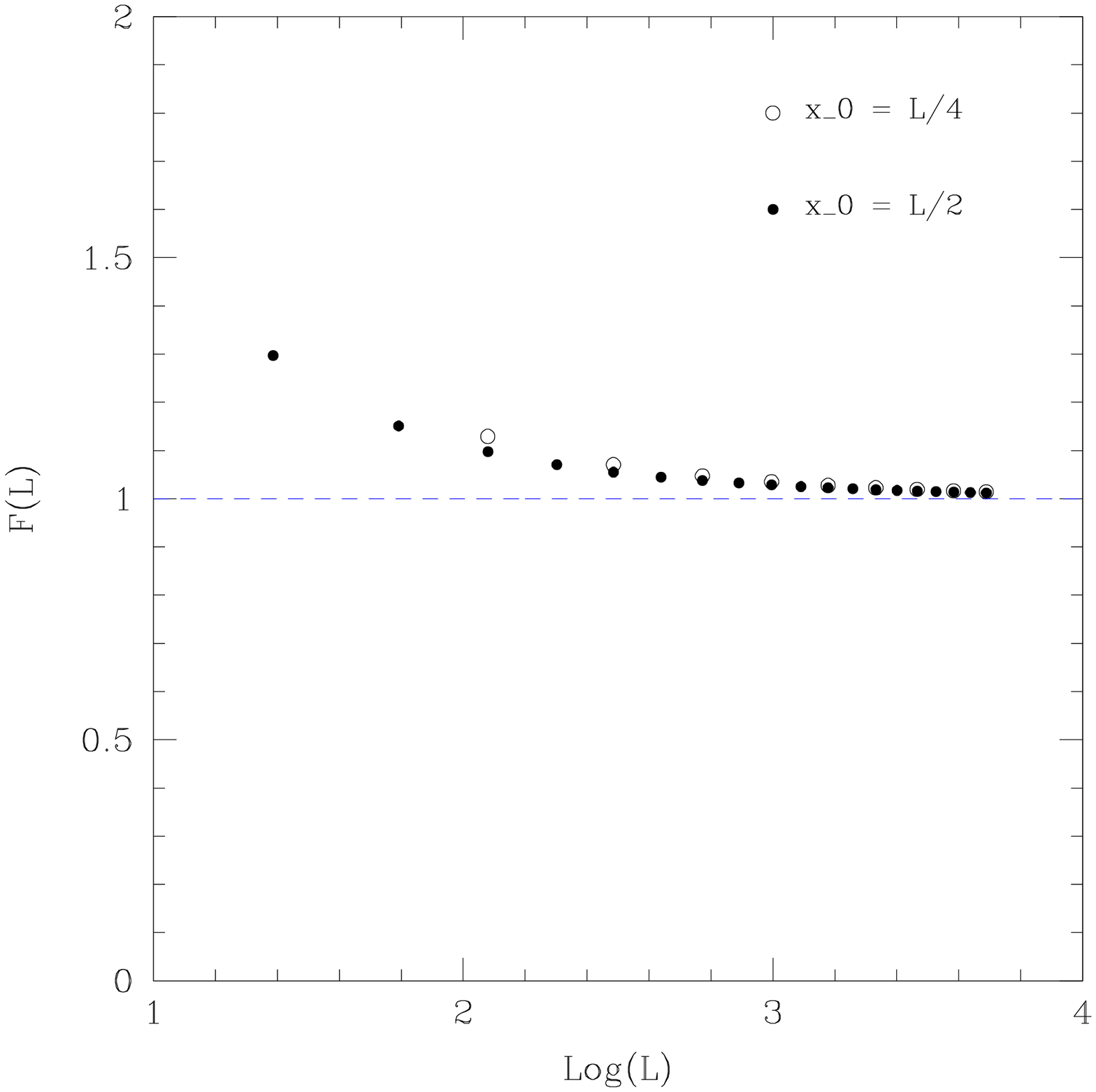,height=4.3in,width=4.5in}}
\vspace{-0.0cm}
\caption{\footnotesize{
The ratio of perturbative finite lattice 
renormalization constants with their continuum value for the second moment 
in the two cases of $x_0 = L/4 (\circ)$ and $x_0 = L/2 (\bullet)$ using 
``finite size momentum''.}
\label{fig:rate2}}
\end{figure}

The lattice results converge much faster to their continuum limit, 
confirming the momentum
quantization as a major source of lattice artefacts. 
Figures \ref{fig:rate1} and \ref{fig:rate2} show the ratio of
lattice over continuum results also for this case. 

In this case also a global fit without elimination of subleading terms provides a very good
estimate of the value of the constant. 
Our final best estimates are given in Table \ref{tab:tetacost}.

\begin{table}
\begin{center}
\begin{tabular}{|@{\qquad} c @{\qquad}|@{\qquad} c @{\qquad}|
@{\qquad} c @{\qquad}|}
\hline
Moment & Definition & Constant \\
\hline
First & $ x_0 = L/4 $ & $B_0 = -0.0577(1)$  \\
\hline
First & $ x_0 = L/2 $ & $B_0 = -0.4129(1)$  \\
\hline
Second & $ x_0 = L/4 $ & $B_0 = -0.2403(5)$  \\
\hline
Second & $ x_0 = L/2 $ & $B_0 = -0.747(2)$  \\
\hline
\end{tabular}
\caption{\footnotesize{
Values of the constants of the correlation functions $f_2$ and $f_3$ 
in the two definitions with the ``finite size momentum''.}}
\label{tab:tetacost}
\end{center}
\end{table}

The result for the constant of the continuum limit of the
quantity  $f_1$ that subtracts the renormalization constants of the $\zeta$ fields
from that of the operator is

\begin{equation}
B_0 = -0.3590(1).
\label{eq:f1_constant}
\end{equation}

\noindent It has been obtained from fits to the data made available to us by P.Weisz.

\begin{table}
\begin{center}
\begin{tabular}{|@{\qquad} c @{\qquad}|@{\qquad} c @{\qquad}|
@{\qquad} c @{\qquad}|}
\hline
Moment & Definition & Constant \\
\hline
First  & $ x_0 = L/4 (p \neq 0) $ & $B_0 = 0.2635(10)$  \\
\hline
First  & $ x_0 = L/2 (p \neq 0) $ & $B_0 = 0.2762(5)$  \\
\hline
Second  & $ x_0 = L/4 (p \neq 0) $ & $B_0 = 0.1875(20)$  \\
\hline
Second  & $ x_0 = L/2 (p \neq 0) $ & $B_0 = 0.1895(50)$  \\
\hline
\hline
First  & $ x_0 = L/4 (\theta \neq 0) $ & $B_0 = 0.12180(15)$  \\
\hline
First  & $ x_0 = L/2 (\theta \neq 0) $ & $B_0 = -0.23340(15)$  \\
\hline
Second  & $ x_0 = L/4 (\theta \neq 0) $ & $B_0 = -0.06080(55)$  \\
\hline
Second  & $ x_0 = L/2 (\theta \neq 0) $ & $B_0 = -0.5675(20)$  \\
\hline
\end{tabular}
\caption{\footnotesize{
Values of the constants of the operators renormalized
in the two definitions with the real momentum $p$ or the ``finite size momentum'' $\theta$ 
different from zero.}}
\label{tab:summ}
\end{center}
\end{table}

Table \ref{tab:summ} contains a summary of the constants $B_0$ for
the operators, after removing the external legs renormalization, in
the various cases that we have discussed.

We have shown by an explicit one-loop calculation the feasibility of a 
lattice evaluation
of the renormalization constants of the first two non-singlet 
twist-2 operators within the Schr\"odinger functional scheme. 
The definition introduced in this paper is suitable for
a finite-volume recursive scheme that can provide a full numerical
reconstruction of the scale dependence of the renormalization constants, 
even in the non-perturbative regime.
By comparing lattice and continuum perturbative estimates,
we can conclude that lattice momentum quantization in a finite volume 
is one of the most important sources of lattice artefacts. The
modification of fermion boundary conditions already introduced in the 
framework of the Symanzik improvement programme carried out by the ALPHA 
collaboration, introduces of a ``finite-size momentum'', which escapes the 
quantization rule and turns out to substantially  improve the approach to the 
continuum of our lattice results.
Numerical studies with standard lattice momenta and with finite-size momenta 
are under way \cite{in_preparation}.\\

\noindent ACKNOWLEDGEMENTS.

\noindent We have considerably profited from the experience accumulated by
the ALPHA collaboration with the Schr\"odinger functional with fermions.
In particular, we thank  K. Jansen, M. L\"uscher, M. Testa and
P. Weisz for many enlightening discussions and M. L\"uscher for the access to some crucial
internal notes of the ALPHA collaboration.
We thank P. Weisz for providing us with the detailed perturbative results 
for ``$f_1$''.

\newpage

\def\NPB #1 #2 #3 {Nucl.~Phys.~{\bf#1} (#2)\ #3}
\def\NPBproc #1 #2 #3 {Nucl.~Phys.~B (Proc. Suppl.) {\bf#1} (#2)\ #3}
\def\PRD #1 #2 #3 {Phys.~Rev.~{\bf#1} (#2)\ #3}
\def\PLB #1 #2 #3 {Phys.~Lett.~{\bf#1} (#2)\ #3}
\def\PRL #1 #2 #3 {Phys.~Rev.~Lett.~{\bf#1} (#2)\ #3}
\def\PR  #1 #2 #3 {Phys.~Rep.~{\bf#1} (#2)\ #3}

\def\etal{{\it et al.}}
\def\ibid{{\it ibid}.}

\end{document}